\documentclass[aps,twocolumn,prb,preprintnumbers,amsmath,amssymb,superscriptaddress]{revtex4-1}
\usepackage{graphicx}
\usepackage{hyperref}
\usepackage{epstopdf}
\usepackage{color}
\usepackage{siunitx}
\usepackage{float}
\usepackage{amsmath}
\usepackage{mathrsfs} 
\usepackage[paperwidth=8.5in,paperheight=11in,centering,hmargin=0.8in,vmargin=0.8in]{geometry}
\begin{document}

\title{Effect of radiation-induced defects on the superfluid density and optical conductivity of overdoped La$_{2-x}$Sr$_x$CuO$_4$}

\author{Fahad Mahmood}
\affiliation{Department of Physics, University of Illinois at Urbana-Champaign, Urbana, 61801 IL, USA}
\affiliation{F. Seitz Materials Research Laboratory, University of Illinois at Urbana-Champaign, Urbana, 61801 IL, USA.}
\affiliation{The Institute for Quantum Matter, Department of Physics and Astronomy, The Johns Hopkins University, Baltimore, MD 21218 USA.}

\author{David Ingram}
\affiliation{Nanoscale and Quantum Phenomena Institute, Department of Physics and Astronomy, Ohio University, Athens, OH 45701, USA.}
 
\author{Xi He}
\affiliation{Brookhaven National Laboratory, Upton, NY 11973, USA.}
\affiliation{Department of Chemistry, Yale University, New Haven, Connecticut 06520, USA.}

\author{ J. A. Clayhold}
\affiliation{SRI International, 201 Washington Road, Princeton, NJ 08540, USA.} 

\author{Ivan Bo{\v{z}}ovi{\'{c}} }
\affiliation{Brookhaven National Laboratory, Upton, NY 11973, USA.}
\affiliation{Department of Chemistry, Yale University, New Haven, Connecticut 06520, USA.}

\author{N.~P.~Armitage}
\affiliation{The Institute for Quantum Matter, Department of Physics and Astronomy, The Johns Hopkins University, Baltimore, MD 21218 USA.}
\affiliation{Canadian Institute for Advanced Research, Toronto, Ontario M5G 1Z8, Canada.}

\date{\today}

\begin{abstract}

Using a combination of time-domain THz spectroscopy (TDTS) and mutual inductance measurements, we have investigated the low-energy electrodynamic response of overdoped La$_{2-x}$Sr$_x$CuO$_4$ films that have been exposed to ion irradiation.  Films went through three rounds of irradiation (2, 4, and 6 $\times 10^{13}$ ions/cm$^2$) and mutual inductance and TDTS experiments were performed between each step.  Together with the as-grown film, this gives four different levels of disorder.   The transport scattering rate that is measured directly in the THz experiments is an approximately linear function of the radiation dose at all temperatures.   This is consistent with a proportionate increase in elastic scattering.  In the superconducting state we find that the relation between $T_c$, the superfluid density, and the scattering rates are quantitatively at odds with the predictions based on the extant theory of Abrikosov-Gorkov-like pair breaking in a dirty $d$-wave superconductor.  Increasing disorder causes only a small change in the superconducting transition temperature for the overdoped films, but the changes to the $\omega \sim 0$ superfluid density are much larger.

\end{abstract}

\maketitle

\section{Introduction}
The normal state of the superconducting cuprates continues to be enigmatic despite more than three decades of intense experimental and theoretical studies. For underdoped cuprates, various forms of structural, magnetic and electronic phases exist above the superconducting transition temperature $T_c$, some of  which seem to be intertwined with superconductivity. On the overdoped side however, the observation of a large and well-defined Fermi surface in both photoemission~\cite{Plate_PRL_2005} and quantum oscillation experiments~\cite{Vignolle_Nat_2008} suggested that the normal state may be a Fermi liquid with small residual interactions and so the superconducting state in turn may be described in terms of conventional Bardeen-Cooper-Schrieffer (BCS)-like physics (albeit one of $d$-wave symmetry). Other experiments, however, indicated anomalies.   For instance, the superfluid density is much lower than expected in this scenario~\cite{Uemura_Nat_1993,Niedermayer_PRL_1993,Panagopoulos_PRB_2003,Corson_PhysicaB_2000,Lemberger_PRB_2010,Lemberger_PRB_2011}.

Recently, a comprehensive study of overdoped La$_{2-x}$Sr$_x$CuO$_4$ (LSCO) films have indicated strong deviations from behavior expected of a BCS $d$-wave superconductor condensing out of a Fermi liquid~\cite{Bozovic_Nat_2016,Mahmood_PRL_2019}. Key observations include an anomalously small superfluid stiffness~\cite{Bozovic_Nat_2016} despite the linear dependence of superfluid density on temperature (the latter usually taken to be a signature of a clean $d$-wave superconductor).  In a previous study, we combined time-domain THz spectroscopy (TDTS) with kHz range mutual inductance (MI) measurements to track both the condensate and the free carrier spectral weight as a function of doping for overdoped La$_{2-x}$Sr$_{x}$CuO$_{4}$ films~\cite{Mahmood_PRL_2019}. We found that a significant fraction of the total spectral weight remained uncondensed as $T\rightarrow0$ and manifests as a Drude-like peak at frequencies where one would expect a superconducting gap.  Observations consistent with our THz ones in overdoped La$_{2-x}$Sr$_x$CuO$_4$ single crystals~\cite{Wang_PRB_76}, found a large fermionic linear-in-T contribution to the heat capacity deep into the superconducting state. In overdoped samples with $T_c \sim 20 $ K, the heat capacity coefficient as $T\rightarrow$0 was roughly 70$\%$ of the normal state and reached essentially 100$\%$ in samples with $T_c \sim$ 7 K. 

There are a number of possible explanations for these latter observations, including the ``dirty $d$-wave" scenario~\cite{Hirschfeld_PRL_1993,Lee-Hone_PRB_2017,lee2018optical}, in which pair-breaking scattering due to even non-magnetic impurities smears out the $d$-wave node leading to a finite density of nodal Bogoliubov quasi-particles and a suppression of $T_c$ and the superfluid density.  Such pair breaking has been often treated in the context of the Abrikosov-Gorkov (AG) theory~\cite{abrikosov1960contribution}.  In conventional s-wave superconductors impurities must be paramagnetic to cause pair breaking.   However in $d$-wave superconductors with a strongly anisotropic gap, potential scatterers break Cooper pairs and cause an AG-like $T_c$ suppression.  A substantial density of non-magnetic impurities is a natural possibility in Sr substituted La$_{2-x}$Sr$_x$CuO$_4$.  Thus, such pair-breaking in both the unitary and Born scattering limits applied to the cuprates has a long history~\cite{Hirschfeld_PRL_1993,Hirschfeld_PRB_1993,borkowski1994distinguishing,fehrenbacher1994gap,Hosseini_PRB_1999,Durst_PRB_2000,Broun_PRL_2007}, with the latter considered recently \cite{Lee-Hone_PRB_2017} to explain the suppression in superfluid density in overdoped samples observed in Ref. \onlinecite{Bozovic_Nat_2016}.   It was argued that a sufficient number of weak Born scatterers could give both the small superfluid density, while retaining its linear temperature dependence.  It was subsequently argued~\cite{lee2018optical} that this general picture could also quantitatively explain the THz data and the large residual THz spectral weight that persists down to the lowest temperatures, as well as the large fermionic heat capacity~\cite{lee2020low}.

However, other possible explanations include large inhomogeneity i.e., the presence of superconducting regions embedded in a normal state metallic background. This scenario is consistent with our observation, from THz measurements, that the width of the T $\rightarrow$ 0 residual Drude peak is about the same as the normal state just above $T_c$.  However, as we discuss in Ref. ~\onlinecite{Mahmood_PRL_2019}, if the residual Drude peak were due to such phase separation then the volume fraction corresponding to the normal metallic region would have to be exceedingly large (e.g., nearly 95$\%$ for the film with $T_c$ = 7 K).   Such extreme phase separation would have to be reconciled with the exceedingly uniform $T_c$ for all the films measured (as characterized by a sharp transition in the dissipative part of the MI).  Recent theoretical work shows that such perspectives could be reconciled if the coherence length of the system is comparable to the correlation length of the disorder potential~\cite{Li2020}.  Moreover, the transition could be more homogeneous than the low-temperature state, if it were governed by a diverging correlation length that averages disorder configurations.  We noted in earlier work~\cite{Mahmood_PRL_2019} that the frequency dependence of the spectral weight distribution was consistent with a scenario of  significant quantum phase fluctuations.

A natural way to test these possibilities and clarify the role of defects is through a systematic investigation of the dependence of the optical conductivity and superfluid density on disorder.  In the present work, we do this by irradiating sequentially two different overdoped LSCO films with 1 MeV oxygen ions as described below. After each round of irradiation, both MI and TDTS experiments are carried out to determine the superfluid density and the real and imaginary parts of the complex conductivity across a broad frequency range.  The low-frequency Lorentzian-like lineshape broadens systematically with increasing fluence.  Its width increases linearly with fluence in a manner consistent with increased disorder scattering.  Increasing disorder leads to a large decrease in the superfluid density.   However, $T_c$ is suppressed much slower than predicted within the AG pair-breaking theory for a $d$-wave superconductor.  The observed relations between $T_c$, the superfluid density, and the scattering rates are incompatible with the AG pair-breaking model. 

\section{Experimental methods}
Complex conductivity was determined by TDTS. In this method~\cite{Mahmood_PRL_2019}, a femtosecond laser pulse is split along two paths and sequentially excites a pair of photoconductive `Auston'-switch antennae grown on low-temperate GaAs wafers. The emitter photoconductive switch is dc biased.   The laser pulse excites charge carriers in the GaAs and ``closes" the switch.  The acceleration of charge on ps time scales creates an almost single pulse of radiation that propagates into free space, is transmitted through the LSCO film, and then measured at the other antenna.  The detection process is roughly the inverse of the generation process, except that the switch bias is provided by the time-varying THz electric field.   The other split-off laser pulse gates the switch.  By varying the length-difference of the two laser paths, one can trace out the electric field of the transmitted pulse as a function of time. Comparing the Fourier transform of the transmission through a LSCO film grown on substrate to that of a bare substrate resolves the full complex transmission. We then invert the transmission to obtain the complex conductivity $\tilde{\sigma}(\nu)$ via the standard formula for thin films on a substrate:  $\tilde{T}(\nu)=\frac{1+n}{1+n+Z_0\tilde{\sigma}(\nu)d} e^{i\Phi_s}$ where $\Phi_s$ is the phase accumulated from the small difference in thickness between the sample and reference substrates, $d$ is the film thickness, $Z_0$ is the impedance of free space ($\approx$377 Ohms), and $n$ is the substrate index of refraction.  Because one measures a complex transmission function, the inversion to complex conductivity is done directly and does not require a Kramers-Kronig transformation.

\begin{figure*}
	\includegraphics[width=0.8\linewidth]{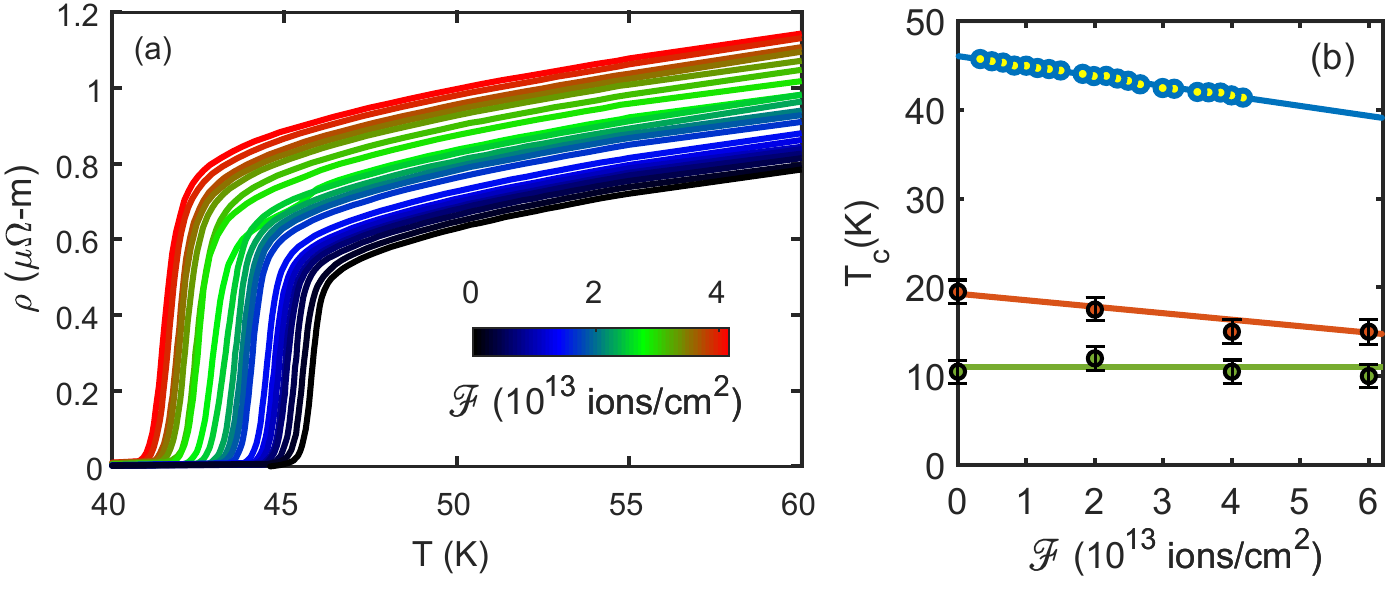}
	\caption{\textbf{(a)} Resistivity $\rho$ as a function of temperature of an optimally doped La$_{1.84}$Sr$_{0.16}$CuO$_4$ on film that was patterned to yield 30 devices.   The film was irradiation with 1 MeV oxygen ions, but with a gradient to the flux such that total fluences acrossess the devices varied between $0$ and $4 \times 10^{13}$ ions/cm$^2$. \textbf{(b)} $T_c$ as a function of irradiation fluence $\mathscr{F}$ for an optimally doped film (blue dots) and the two overdoped films studied in this work (red and green dots). Solid lines are guides to the eye.}
	\label{Fig1}
\end{figure*}

When a sample goes superconducting, spectral weight is  transferred from the conducting response of the normal state into a zero-frequency delta function.  The superfluid spectral weight in the delta function $S_{\delta}$ is determined by a two-coil mutual inductance (MI) technique. It provides a measure of the complex conductivity at 40 kHz. This is six orders of magnitude lower frequency than the THz range and can be considered effectively the dc limit.  For a uniform superconducting film of thickness \textit{d} and infinite radius placed between two coils of radii $R_1$ and $R_2$ parallel to one another and separated by a distance $D$, the complex mutual inductance can be written as:
\begin{align*}
\hat{M} &= \mathrm{Re}M +i\mathrm{Im}M  \\
& = \mu_0\pi R_1R_2\frac{\int_{0}^{\infty}d\mathbf{q} [\mathrm{exp}(-\mathbf{q}D)J_1(\mathbf{q}R_1)J_1(\mathbf{q}R_2)]}{\mathrm{cosh}(Qd) + [(Q^2+\mathbf{q}^2)/2\mathbf{q}Q]\mathrm{sinh}(Qd)}
	\end{align*}
\noindent where $\mathbf{q}$ is the wave-vector, $J_1(x)$ is the first order Bessel function, $Q^2 = q^2 + (1/\lambda^2) -i\mu_0\omega\sigma_1 $ and $\sigma_2 = 1/\mu_0\omega\lambda^2$. This can be generalized to the case of two solenoids with $N_1$ and $N_2$ turns respectively by the summation over pairs of coils. Re$M$ and Im$M$ are measured experimentally and the above equation is numerically inverted using the algorithm outlined in Ref. \onlinecite{He_RSI_2006} to obtain $\lambda$ and $\sigma_1$ from which the superfluid spectral weight is calculated as $S_{\delta} =  \frac{1}{2\pi\mu_0\lambda^2}$. Further details on the mutual inductance setup can be obtained in the Methods section of Ref. \onlinecite{Bozovic_Nat_2016} and further details on the comparison of MI to TDTS can be found in Ref. \onlinecite{Mahmood_PRL_2019}.

\section{Materials Preparation}
The LSCO films were deposited by atomic-layer-by-layer molecular-beam-epitaxy (ALL-MBE) on $10 \times 10 \times 1$-mm$^3$ single-crystal LaSrAlO$_4$ substrates in ozone partial pressure.  Substrates were epitaxially polished perpendicular to the (001) direction.   A buffer layer was grown that is effectively one monolayer of La$_3$AlCuO$_7$.  Overdoped LSCO is prone to formation of oxygen vacancies, the concentration of which increases nonlinearly with overdoping~\cite{radaelli1994structural}.  MBE growth requires high vacuum, but once films are grown they are post-annealed in situ, improving the film crystallinity and making the superconducting transition sharper.  Overdoped LSCO films are annealed under the 10$^{-4}$ Torr ozone partial pressure at 600$^\circ$C for up to 4 hours and then cooled down under the same ozone environment to room temperature~\cite{leng2015controlling}.  Films were characterized by reflection high-energy electron diffraction, atomic force microscopy, X-ray diffraction, resistivity, and magnetization measurements, all of which indicate excellent film quality. All the films studied in this work are 20 monolayers thick (one monolayer is 0.66 nm).  Two overdoped LSCO films with as-grown $T_{c,0} = 19.5$ and $T_{c,0} = 10.5$ K were studied with mutual inductance and TDTS. Unlike crystals LSCO films can have charge densities different than set by their Sr levels due to oxygen non-stochiometry.  We use the calibration of Ref.~\cite{presland1991general} to set the two doping levels of films A and B to $p=0.239$ and $0.255$. A third film, close to optimally doped, was irradiated and investigated primarily with dc resistance measurements.

Ion irradiation was performed at Ohio University in the Edwards Accelerator Laboratory, using the 4.5 MV tandem accelerator.   Oxygen ions at 1 MeV were used.  In this energy range damage rates increase at lower energies, but using even lower energies would increase the necessary irradiation time, as the beam intensity decreases below 1 MeV.  Stopping and Range of Ions in Matter (SRIM)~\cite{ziegler1985stopping,SRIM} calculations indicate that the main effect on the 13.2 nm thick film of La$_{1.84}$Sr$_{0.16}$CuO$_4$ is the formation of columnar damage tracks. The energy was chosen for a stopping range much greater than the film thickness i.e. so that the largest part of the incident energy is deposited deep in the substrate below the film.   Ions in this energy range and fluence do not oblate the material.  From previous modeling and experiments, it is expected that narrow ($\sim$0.4 nm) columnar defects tracks are produced throughout the film~\cite{clayhold2010constraints}.  With an in-plane lattice constant of 0.38 nm in LSCO and a Fermi wavelength of approximately twice this, these defects can be regarded as being almost line-like scatterers.  Note that these ions are in a completely different parameter regime than the very high energy ($\sim$4 GeV ions) Au ions used to make columnar defects and enhance vortex pinning in cuprates~\cite{petrean2001thermal}.   Irradiation with such high energy ions results in defects with diameters almost 50 times larger.  Oxygen ions at 1 MeV energy also have the advantage over proton or electron irradiation, which tend to create point defects that self-anneal at room temperature.  Samples with such point defects change their properties over the time scale of a few days to a week~\cite{hughes1975real,petrean2001thermal}.  Columnar defects are less susceptible to self-annealing.  A previous study showed an absence of time-dependence and aging effects~\cite{clayhold2010constraints}.  We performed three rounds of irradiation (2, 4, and 6 $\times 10^{13}$ ions/cm$^2$) and performed mutual inductance and TDTS experiments after each step. With the two films A and B that THz experiments were performed on, four different disorder compositions (including the as-grown) were measured.  A third optimally doped film C was irradiated in a fashion such as to introduce a substantial fluence gradient across it~\cite{clayhold2010constraints}.  This film was patterned using optical photoresist and a combination of ion milling and chemical etching.  More details of the system design, film growth, and patterning have been previously described~\cite{bozovic2001atomic}.  Resistivity measurements were performed on the patterned film C irradiated with a fluence gradient in a specially constructed system, which is capable of making dc resistivity measurements at 30 equally spaced locations on a single film~\cite{clayhold2008combinatorial}.  This gives 30 resistive measurements on 30 different disorder levels from a single film at the ssame time.

\begin{figure*}
	\includegraphics{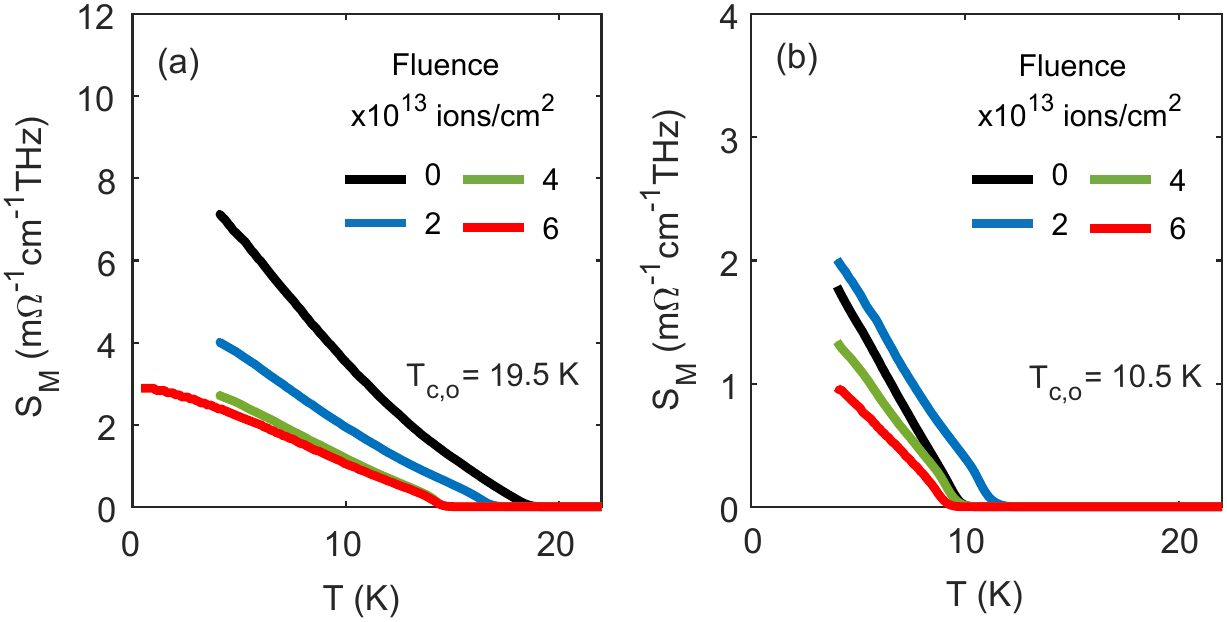}
	\caption{The superfluid spectral weight $S_{\delta}$ as a function of temperature for various total irradiation fluences for two overdoped La$_{2-x}$Sr$_{x}$CuO$_4$ films with original (e.g. pre-irradiation)   \textbf{(a)} $T_{c,0}$=$\SI{19.5}{\kelvin}$ and \textbf{(b)} $T_{c,0}$=$\SI{10.5}{\kelvin}$. $S_{\delta}$ is derived from the complex impedance using a two-coil MI setup ($\nu = \SI{40}{\kilo\hertz}$).
	For the film with $T_{c,0}$= $\SI{19.5}{\kelvin}$, measurements were performed down to $\SI{300}{\milli\kelvin}$ in a He$^3$ system after irradiation with a total fluence of $\SI[per-mode=symbol]{6e+13}{ions\per\centi\meter\square}$ (red line in \textbf{(a)}).} 
	\label{Fig2}
\end{figure*}

\begin{figure*}
	\includegraphics{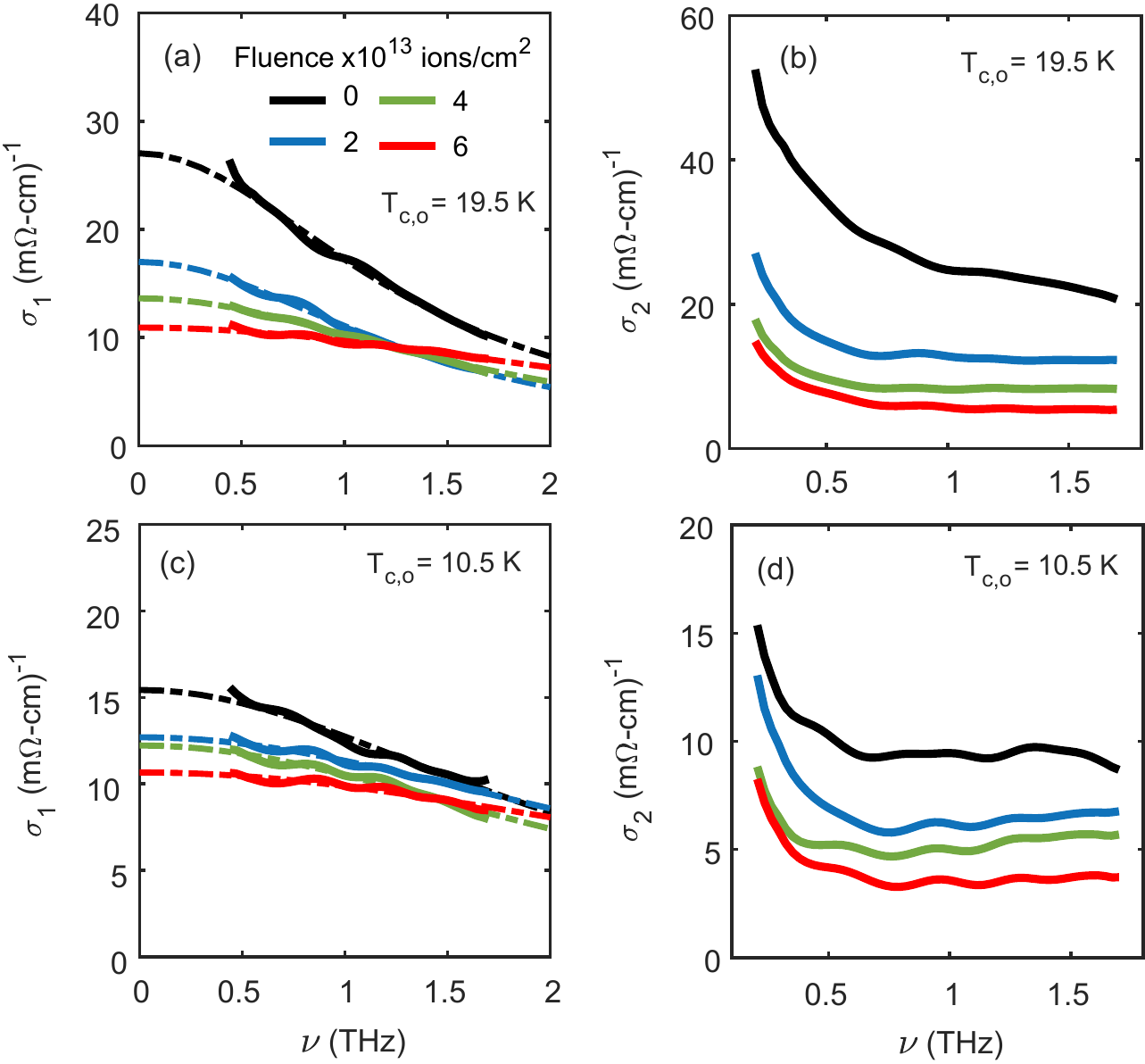}
	\caption{Real ($\sigma_1$) and imaginary ($\sigma_2$) parts of the THz optical conductivity as a function of frequency ($\nu$) after irradiation with a total fluence between zero and $\SI[per-mode=symbol]{6e+13}{ions\per\centi\meter\square}$. \textbf{(a)} and \textbf{(b)} show the data for the film that had $T_{c,0}$ of $\SI{19.5}{\kelvin}$.  \textbf{(c)} and \textbf{(d)} show the data for the film that had $T_{c,0}$ of $\SI{10.5}{\kelvin}$. All data are taken at a temperature of $T=\SI{1.6}{\kelvin}$. Dashed lines in \textbf{(a)} and \textbf{(c)} are fits to a single Drude from as described in the text.}
	\label{Fig1}
\end{figure*}

\section{Data and analysis}

Fig. 1 shows the resistivity $\rho$(T) of the optimally doped La$_{2-x}$Sr$_x$CuO$_4$ film. C (x = 0.16) as a function of temperature $T$ after irradiation with increasing doses of 1 MeV oxygen ions. The fluence of irradiation is varied from 0 to $\SI[per-mode=symbol]{4e+13}{ions\per\centi\meter\square}$. With increasing fluence, the transition temperature $T_c$ of the film decreases continuously (as also shown in Fig.~1b). Crucially, the slope of the temperature dependence of the $\rho(T)$ curves in the normal state does not change much with increasing fluence, while the residual resistivity increases with increasing fluence and shifts the curves upward.  Previous experiments using these ions~\cite{clayhold2010constraints} have shown that the residual resistivity of these films is a linear function of the radiation dose at all temperatures.  This indicates that the additional disorder neither traps holes nor dopes additional carriers.  It also indicates that insulating columnar defects are not being created~\cite{valles1989ion} as they would change the resistivity by a fluence-dependent multiplicative factor. These data indicate that we could regard the effects of disorder as changing the disorder's contribution to resistivity in a fashion expected from Boltzmann transport.

To study the effects of ion irradiation on the superfluid density of the two $T_{c,0} = 19.5$ and $T_{c,0} = 10.5$ K overdoped LSCO films, we performed mutual inductance experiments on the films initially and after each round of irradiation during which the film is irradiated with an additional fluence of 2 $\times 10^{13}$ ions/cm$^2$. Fig. 2 shows the extracted superfluid spectral weight ($S_\delta$) from these measurements after each round. As can be seen, the superfluid density remains linear as a function of temperature down to at least $\SI{4}{\kelvin}$ after 3 rounds of irradiation (total of 6 $\times 10^{13}$ ions/cm$^2$).  For all films, there is either small or no decrease in $T_c$ with increasing irradiation as determined by the temperature where $S_{\delta} \rightarrow 0$, whereas the changes to the superfluid density are much larger.  $T_c$ as a function of irradiation fluence is plotted in Fig.~1b for all three films. For film A with $T_{c,0} = \SI{19.5}{\kelvin}$, $T_c$ decreased by 4.5 K  whereas for the B film with $T_{c,0} = \SI{10.5}{\kelvin}$, the change in $T_c$ is less than 1 K after a total fluence of 6 $\times 10^{13}$ ions/cm$^2$. These results highlight that increasing disorder causes only a small change in $T_c$ for the overdoped films, but the changes to the superfluid density are much larger.

\begin{figure*}
	\includegraphics{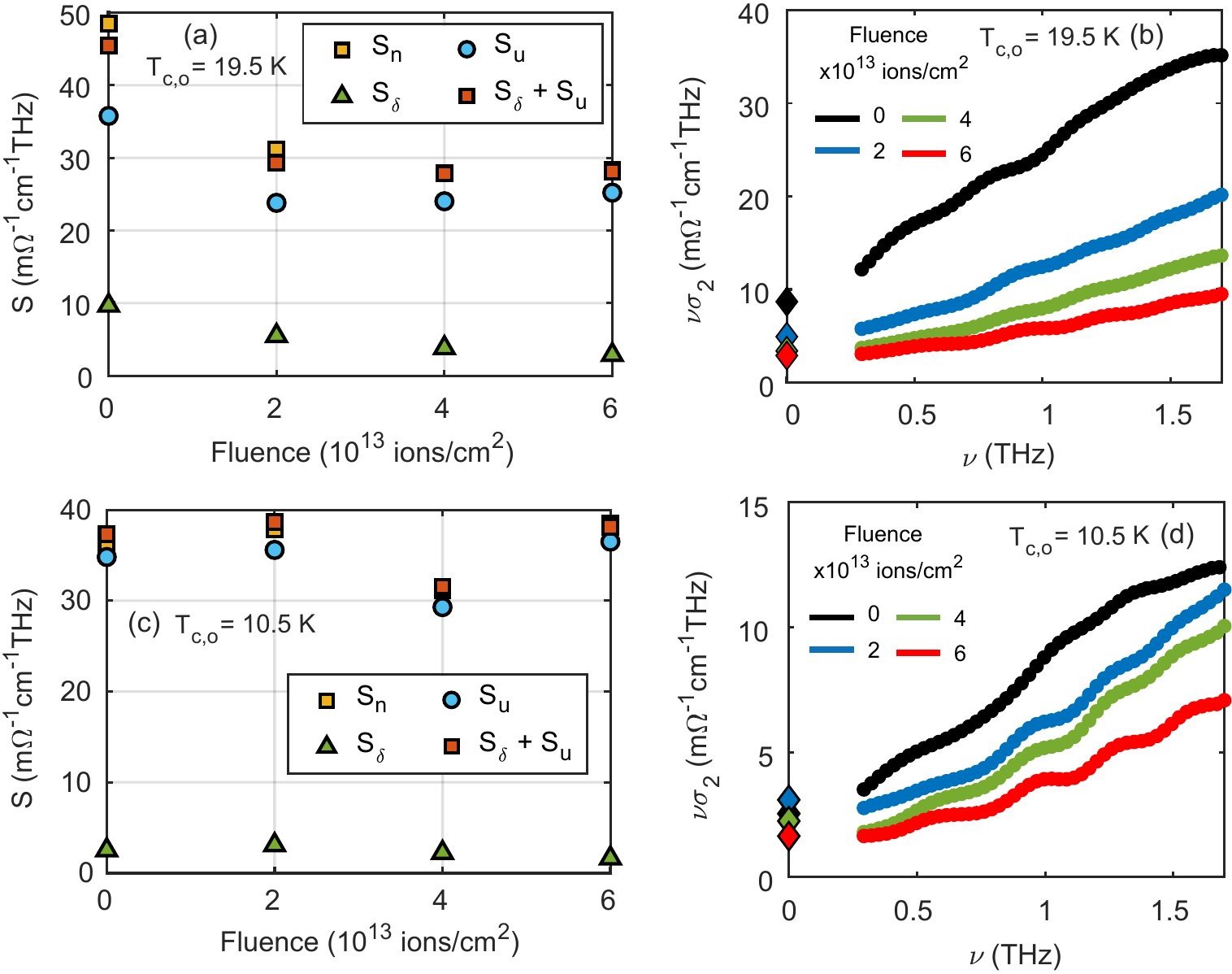}
	\caption{ \textbf{(a)} Spectral weight of the various conductivity contributions: the normal state ($S_{n}$), superfluid ($S_{\delta}$), uncondensed carriers ($S_{u}$), and the sum $S_{\delta}$ and $S_{u}$  after various irradiation fluences at $T = \SI{1.6}{\kelvin}$ for the film with $T_{c,0}$ of $\SI{19.5}{\kelvin}$.	$S_{\delta}$ is determined from the MI data shown in Fig.~2 while $S_{u}$ and $S_{n}$ are determined from Drude fits to $\sigma_1(\nu)$.  \textbf{(b)} $\nu\sigma_2$ versus frequency at $T = \SI{1.6}{\kelvin}$ after various irradiations for the film with $T_{c,0}$ of $\SI{19.5}{\kelvin}$. Circle and diamond symbols represent the TDTS and MI data respectively. \textbf{(c)} and \textbf{(d)} Same as \textbf{(a)} and \textbf{(b)} but for the film with $T_{c,0}$ of $\SI{10.5}{\kelvin}$.}
	\label{Fig1}
\end{figure*}

Fig. 3 shows the measured real and imaginary parts of the THz-range optical conductivity ($\sigma_1$ and $\sigma_2$) of the two overdoped La$_{2-x}$Sr$_x$CuO$_4$ films before and after irradiation. For both films, regardless of disorder, $\sigma_1$ exhibits a large residual ``normal" peak at low temperature. This was noted previously for unirradiated overdoped LSCO films~\cite{Mahmood_PRL_2019} as well as for Bi2212 films~\cite{Corson_PhysicaB_2000}.  With increasing disorder this peak becomes broader.  Additionally, $\sigma_2$ shows a strong upturn at low frequencies indicative of superconductivity.  Regardless of disorder levels, the real part of the conductivity can be fit to a single Drude-like peak i.e., $\sigma_1(\nu) = \frac{S/\gamma}{1+\nu^2/ \gamma^2}$ (dashed lines on Fig. 3a and 3c). We fit all the $\sigma_1$ data to this Drude form at all temperatures to extract out the effective scattering rate $\gamma$ and spectral weight $S$.   Fitting to just the real part was the same method we used in Ref.~\cite{Mahmood_PRL_2019}.   We showed theirin that the errors introduced by not fitting to $\sigma_1$ and $\sigma_2$ simultaneously were small.  In the normal state we denote the spectral weight $S_n$ and the uncondensed spectral weight in the superconducting state in the superconducting state we denote $S_u$.

In Figs.~4a and 4c we show the spectral weights of the various conductivity components as a function of radiation fluence.  We plot the spectral weight of the normal state conductivity peak ($S_{n}$), superfluid delta function ($S_{\delta}$),  the residual uncondensed carriers ($S_{u}$), and the sum of $S_{\delta}$ and $S_{u}$.  Here we note that, similar to the results in Ref.~\onlinecite{Mahmood_PRL_2019}, the Ferrell-Glover-Tinkham (FGT) sum rule for conservation of the optical spectral weight is satisfied regardless of irradiation disorder~\cite{ferrell1958conductivity,tinkham1959determination}.  One can see that despite the increasing proportion of the low temperature uncondensed spectral weight the sum of the spectral weights $S_{\delta} + S_{u}$ equals  $S_{n}$ to within experimental error.  For the $T_{c,0} = \SI{19.5}{\kelvin}$ sample there is only a small decrease of the spectral weight with increasing fluence. 

To further check these dependencies, we plot the quantity $\nu\sigma_2$ (proportional to the superfluid density) as a function of $\nu$ as determined from TDTS experiments after each round of radiation (Fig.~4b and 4d). On the same plot, we also plot $S_{\delta}$ as obtained from MI experiments at 40 kHz.  As 40 kHz is much smaller that the THz scale, this shows that for all cases, $\displaystyle{\lim_{\nu \to 0}} \; \nu\sigma_2$ is roughly equal to $S_{\delta}$. This provides an additional consistency check that our measured $\sigma_1$ and $\sigma_2$ are Kramers-Kronig consistent and confirms the reliability of single-Drude fits in Fig.~3 to extract the spectral weight and scattering rates.   Note that the quantity  $  \nu\sigma_2 $ can be interpreted as a frequency (and length) dependent superfluid stiffness~\cite{Bilbro_Natphys_2011}.   Within this interpretation our data indicates that the superfluid stiffness is primarily degraded on the longest length scales (also seen in the fact that $S_{\delta}$ changes a lot with increasing disorder, but the full spectral weight in the superconducting state $S_{\delta} + S_{u}$ does not).

Fig. 5a shows the extracted scattering rate $\gamma$ with temperature for the film with  $T_{c,0} = \SI{19.5}{\kelvin}$ after each round of irradiation with a total fluence from  $0$ to $\SI[per-mode=symbol]{6e+13}{ions\per\centi\meter\square}$. As shown in Fig.~5, for each of the four cases, $\gamma$ monotonically increases above $T_c$ and the scattering rate of the $T\rightarrow 0$ residual Drude is about the same as that in the normal state right above $T_c$. More importantly, there is a substantial increase in $\gamma$ with increasing irradiation. 

Our scattering rate results are summarized in Fig.~5b where we note the value of $\gamma$ (in units Kelvin) of the residual Drude peak widths (of both samples) after each round of irradiation versus irradiation fluence. For the $T_{c,0} = \SI{19.5}{\kelvin}$ film, the scattering rate $\gamma$ after irradiation with a fluence of  $\SI[per-mode=symbol]{6e+13}{ions\per\centi\meter\square}$ is roughly 2.5 times than that of the non-irradiated film. In contrast, as discussed above, $T_c$ changes only modestly from $\SI{19.5}{\kelvin}$ to $\SI{15}{\kelvin}$ ($\sim 23\% $).  The superfluid density extrapolated to low temperature changes by approximately a factor of 3 (e.g. close to the scale of the scattering rate change). For the film with $T_{c,0} = \SI{10.5}{\kelvin}$, the scattering rate $\gamma$ after irradiation with a fluence of  $\SI[per-mode=symbol]{6e+13}{ions\per\centi\meter\square}$ is roughly 1.35 times that of the non-irradiated film yet the overall change in $T_c$ is negligible.  Here again the superfluid density of the extrapolated low temperature superfluid density is changed by approximately the same factor as the scattering rate i.e. approximately a factor of 1.4.   $T_c$ is plotted as a function of the scattering rate $\gamma$ in Fig.~5c.

\begin{figure*}
	\includegraphics[width=17.5cm]{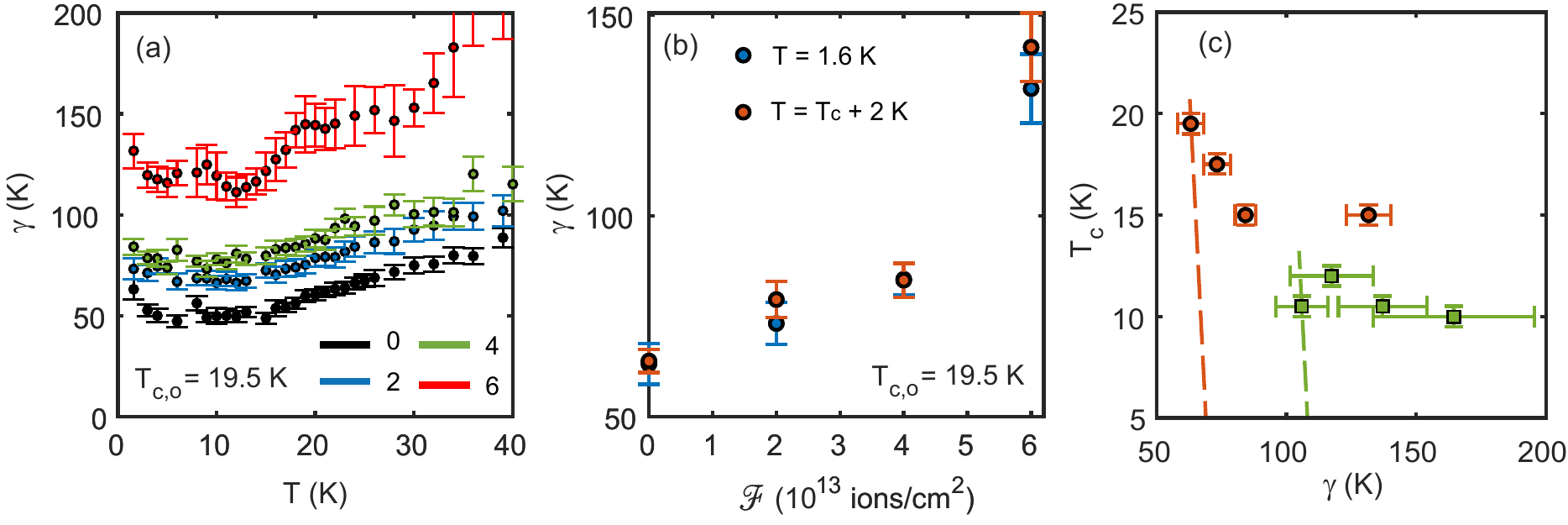}
	\caption{ \textbf{(a)} Temperature dependence of the scattering rate $\gamma$, in units Kelvin, with temperature for the film with $T_{c,0} = \SI{19.5}{\kelvin}$ after various irradiation fluences. $\gamma$ is obtained with a single Drude-fit to $\sigma_1(\nu)$ at all temperatures. \textbf{(b)} $\gamma$ as a function of irradiation fluence for the film with $T_{c,0} = \SI{19.5}{\kelvin}$ at $T=\SI{1.6}{\kelvin}$ and at $\SI{2}{\kelvin}$ above $T_c$.   \textbf{(c)}  $T_c$ vs. scattering rate for both samples.  Dashed lines are the predicted dependence of $T_c$ on the scattering rate for the Abrikosov-Gorkov pair breaking theory as described in the text from the linearized theory that predicts $dT_c/d \gamma = - \pi^2/4$ for small scattering~\cite{borkowski1994distinguishing,fehrenbacher1994gap}.   The plotted lines here are constrained to pass through the zero fluence samples.}
	\label{Fig5}
\end{figure*}

\section{Discussion}

The above experimental observations provides an opportunity to test extant theories of the effects of disorder on $d$-wave superconductors.  For instance, at the largest disorder, the superfluid density of the $T_{c,0}$=19.5  K sample decreases by a factor of approximately 2.5, when the scattering rate increases by a factor of 3, and the $T_{c,0}$=10.5 sample's superfluid density decreases by a factor of 2, when the scattering rate increases by a factor of 2.   This is as expected for a model where the superconducting gap scales with $T_c$ as the FGT sum rule~\cite{ferrell1958conductivity,tinkham1959determination} predicts the superfluid density $\rho \propto \Delta$.

With regards to the normal state transport, the results in Fig.~1a and Fig.~5c taken together signifies that irradiation gives a contribution to disorder scattering that is linear in fluence, and that we can regard the effects of disorder in a fashion expected from Boltzmann transport theory.  Although changes in $T_c$ are modest, at least for the optimally doped sample the changes in $T_c$ vs. irradiation dose and the changes to the residual resistivity are very close to what was found in previous studies with for instance 2.5 MeV electron irradiation~\cite{rullier2003influence} (which phenomenologically seems to introduce similar effects to oxygen ions at our energies).

The idea that $T_c$ is driven to small values (particularly in overdamped samples) from a much higher hypothetical ``clean" $T_{c0}$ is one with much history, both older~\cite{kresin2006inhomogeneous,kresin1996effect} and recent~\cite{pelc2019unusual,pelc2018emergence}.   There are number of ways to treat disorder in systems like the cuprates.  Our observations appear to be inconsistent with theoretical models mentioned above that invoke dirty BCS $d$-wave superconductivity to explain the unusually low superfluid phase stiffness in La$_{2-x}$Sr$_x$CuO$_4$ and treat disorder in terms of an effective medium approximation, in which fluctuations in the disorder landscape are averaged out, and the superconducting state is homogeneous~\cite{Hirschfeld_PRL_1993,Lee-Hone_PRB_2017,lee2018optical}. In such models, starting from an AG-like model of pair breaking~\cite{abrikosov1960contribution}, it was argued that the superfluid density and THz conductivity of overdoped LSCO are compatible with a Landau Fermi liquid/BCS description of these samples, provided dopants are treated within the ``dirty $d$-wave " BCS theory and assuming primarily weak Born-like scatterers. We note that while the calculations showed good agreement with some previous data, they relied on three key assumptions: 1) a large concentration of almost infinitely weak scatterers 2) a clean limit superconducting transition temperature $T_c$ ($T_{c0}$) that is much higher than what has actually been observed and 3) an actual $T_c$ that is set by disorder.  Within this picture, the superconducting $T_c$ is strongly suppressed by disorder.  As is clear from Figs.~1b and 5c, the third assumption is clearly not valid. 

Within the AG pair breaking theory one expects a very strong dependence of $T_c$ on the scattering rate.  The linearized theory (valid for small scattering) predicts $dT_c/d \gamma = - \pi^2/4$~\cite{borkowski1994distinguishing,fehrenbacher1994gap}.   The dashed lines plotted in Fig. 5c are the expected dependence of $T_c$ on the scattering rate for the Abrikosov-Gorkov pair breaking theory.   One can see that the experimentally observed dependence is far weaker than the AG theory predicts.  

Putting it differently, the authors of Refs.~\onlinecite{Lee-Hone_PRB_2017,lee2018optical} quantify the phenomenology of overdoped cuprates by using the ratio $\gamma/T_{c0}$ where $\gamma$ is the normal-state scattering rate and $T_{c0}$ is the  transition temperature of a hypothetical corresponding clean system. In Refs.~\onlinecite{Lee-Hone_PRB_2017,lee2018optical}, the decrease in $T_c$ with overdoping is explained by the increase in $\gamma$ with overdoping for a $T_{c0} \sim \SI{80}{\kelvin}$.  However in the present case, a small decrease in $T_c$ with an increase in irradiation disorder is only consistent with a proportionally small change in the scattering rate of the normal and the residual uncondensed carriers. That is, we expect a Drude-like residual conductivity as $T\rightarrow0$ whose width scales as a function of the change in $T_c$. Our data rule this out. In fact, for the dirty $d$-wave BCS theory to explain our observation the clean-limit $T_{c0}$ would have to increase with increasing disorder.  This is clearly against expectations.  For the sample with $T_{c,0} = \SI{19.5}{\kelvin}$, the scattering rate upon the maximal irradiation increases by a factor of 2.5, while $T_c$ drops only by about 25$\%$.

A number of other recent papers have looked at the effect of disorder on $d$-wave superconductivity from other perspectives.  Ref. \onlinecite{Li2020} emphasizes the possibility of granularity that arises from $d$-wave superconductivity in a $t-J$  model and relatively flat bands in the antinodal regions of the Brillouin zone in the presence of homogeneous disorder.  When the superconducting coherence length is comparable to the correlation length of the disorder potential, the pair-field amplitude becomes spatially heterogeneous.  A feature of the inhomogeneous state is a large density of gapless quasi-particle states that arise from the normal metallic regions between the superconducting grains.   These could account for both the uncondensed carriers seen in the THz conductivity and the fermionic heat capacity.   This model also shows a dependence of the mean-field $T_c$ on the scattering rate that is much slower than the AG prediction~\cite{Li2021}, which is in accord with our observation.  Ref. \onlinecite{sulangi2018quasiparticle} argued through detailed analysis and calculation that the amount of random (unitary or Born) impurities necessary to create the fermionic heat capacity is inconsistent with the constraints of scanning tunneling spectroscopy experiments.   They point out that the disorder profile of the cuprates is consistent with smooth disorder due to off-plane impurities.  However, the disorder from ion tracks is expected to be columnar in our experiments.  We hope a number of these approaches can be developed to be able to make explicit comparison to our results.

\section{Conclusion}
We have used a combination of time-domain THz spectroscopy and mutual inductance measurement, to investigate the low-energy electrodynamic response of overdoped La$_{2-x}$Sr$_x$CuO$_4$ films, which were damaged by ion irradiation.  The transport scattering rate (measured directly in the THz experiments) is an approximately linear function of the radiation dose at all temperatures.   We find that the dependence of $T_c$ on scattering rates is qualitatively at odds with the predictions based on the extant theory of Abrikosov-Gorkov-like pair-breaking in a dirty $d$-wave superconductor.

\begin{acknowledgments}

 The authors would like to thank L.~Benfatto, P.~Hirschfeld, S.~Kivelson, Z.~Li, and J. Zaanen for helpful discussions. Research at JHU was funded under the auspices of the Institute for Quantum Matter, an EFRC funded by the DOE BES under DE-SC0019331.  N.P.A. had additional support through the Quantum Materials program at the Canadian Institute for Advanced Research.  Film synthesis by molecular beam epitaxy and characterization was done at BNL and was supported by the U.S. Department of Energy, Basic Energy Sciences, Materials Sciences and Engineering Division. X.H. is supported by the Gordon and Betty Moore Foundation's EPiQS Initiative through Grant GBMF9074 to I.B.  Work by J.A.C. was performed at Miami University in Oxford, OH and at Brookhaven National Laboratory.
 \end{acknowledgments}
 
 \bigskip

\noindent
\textbf{Correspondence:} Correspondence and requests for materials should be addressed to F.M.~(fahad@illinois.edu) or N.P.A.~(npa@pha.jhu.edu).

\bibliography{references} 

\end{document}